# Designing Singing Syllabi with Virtual Avatars: AI-Assisted Syllabus Reauthoring

Xinxing Wu[1]

## Abstract

Traditional syllabi often function as static reference documents rather than engaging introductions to a course. In practical teaching, we observe that few students thoroughly read or fully comprehend the information provided in traditional, text-based course syllabi, which can leave essential information underused. This paper reframes syllabus communication as a design problem and documents an AI-assisted workflow for transforming a traditional syllabus into a musical, video-based, and avatar-enhanced learning artifact. The paper traces the process of lyrical adaptation, music generation, video composition, avatar synthesis, and optional browser-based interaction. And the paper contributes a reproducible workflow and a concrete example of syllabus reauthoring. The discussion in this paper positions the singing syllabus as a supplement to, not a replacement for, the formal written syllabus and identifies future directions for empirical evaluation. The complete implementation described in this paper is publicly available at https://github.com/xinxingwu-uk/SSVA

[1] Kentucky State University, USA

**Corresponding Author:**

Xinxing Wu, School of Mathematics and Computer Science, Kentucky State University

Emails: Xinxing.Wu@kysu.edu; xinxingwu@gmail.com

## 1. Introduction

Modern curricula must evolve beyond traditional subject matter, integrating emerging skills and competencies to remain relevant in a rapidly changing world [Stephens et al., 2008]. An effectively designed curriculum acts as a supportive framework that enriches the educational experience for both instructors and students, fostering greater engagement, clarity, and meaning. The use of a syllabus in higher education is common practice [Doolittle et al., 2010], and it typically serves as a vital foundational element for instructional communication [Gin et al., 2021]. Syllabi typically outline essential details such as instructor and course information, learning outcomes, weekly topics, grading criteria, academic policies, and course expectations, and it collectively forms the structural backbone of the entire course [Harrington et al., 2015]. A course syllabus serves not only as a roadmap for navigating the term and its content but, as researchers suggest, also holds potential to support students' self-directed learning [Bosch et al., 2022; Botha et al., 2025]. Yet, despite their importance, few students dedicate sufficient time to thoroughly reading syllabi or recalling basic syllabus information [Howton et al., 2020; Rumore, 2016]. Most students either discontinue using the syllabus after the initial weeks or ignore it entirely thereafter. Research and faculty observations confirm that students frequently skim syllabi or overlook them entirely, often leading to confusion about assignments, deadlines, and grading policies [Lopez et al., 2024]. This widespread student disengagement with syllabi represents a persistent, yet frequently overlooked, barrier to effective course delivery and learning outcomes.

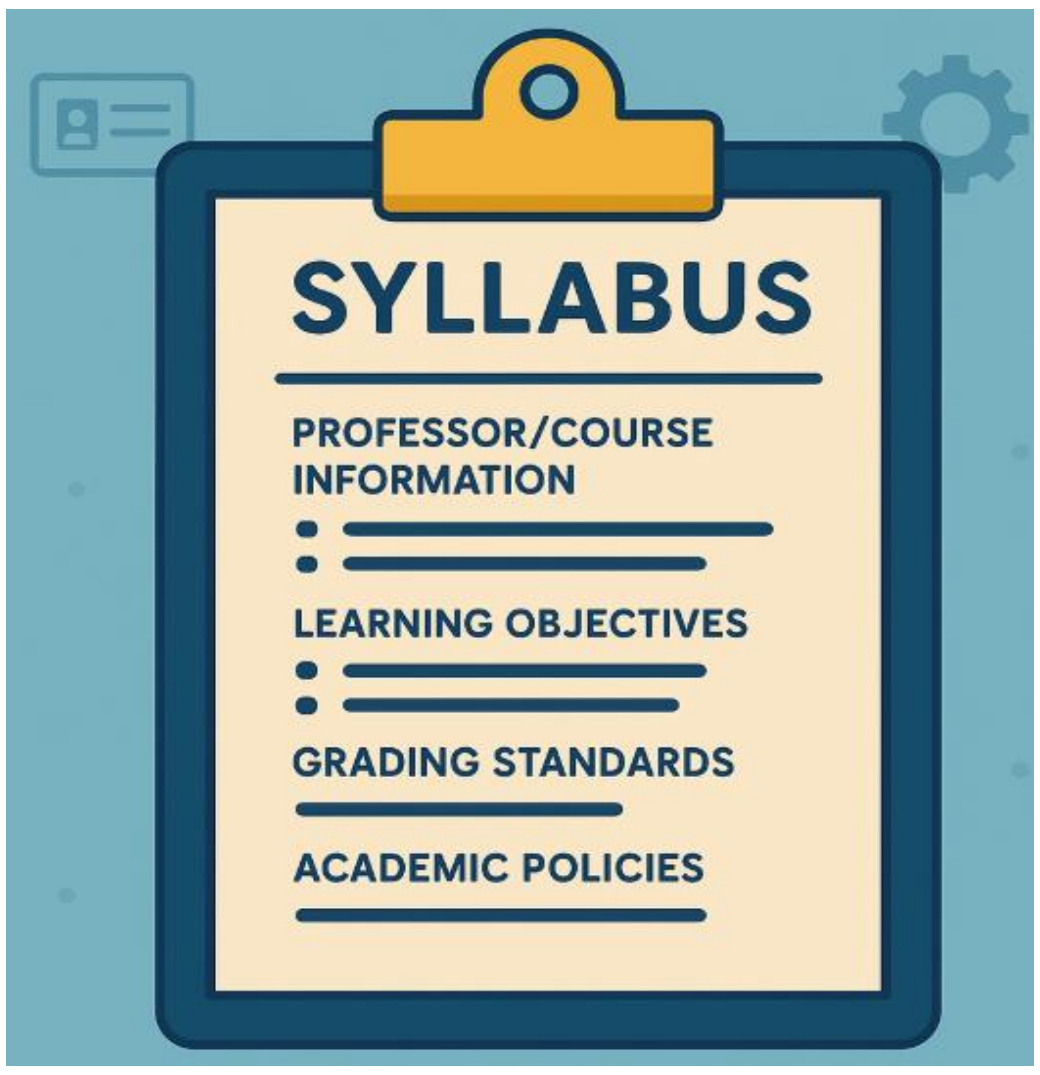

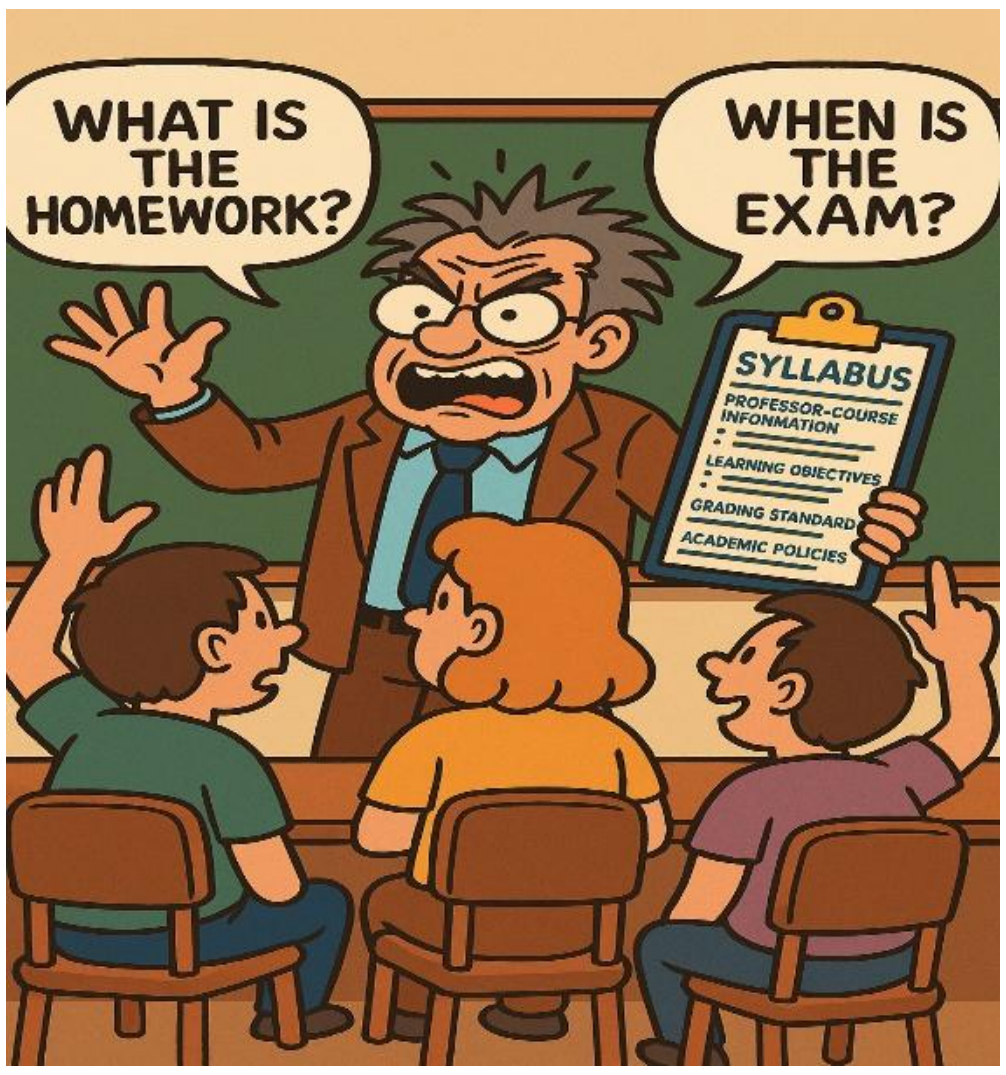

*Figure 1. A humorous cartoon illustrating the traditional text-based syllabus and students' recurring questions (Here, the two images are generated by utilizing ChatGPT[1]).*

In a media environment shaped by platforms such as YouTube Shorts and Spotify, today's students are increasingly accustomed to multimedia content, and research suggests they respond especially well to materials that are brief, engaging, and emotionally designed [Ayres et al., 2025; Navarrete et al., 2025]. Consequently, the traditional syllabus, usually presented as a static, text-heavy document in Word/PDF format or hosted on a Learning

[1] https://chatgpt.com/

Management System (LMS), may be less engaging for many contemporary students, who are increasingly accustomed to dynamic and interactive digital media experiences. This shift compels educators to fundamentally reconsider how they present course materials, aligning them with contemporary attention patterns while maintaining crucial academic rigor [Mayer, 2022; Chan et al., 2023; Thelma et al., 2024].

In this paper, we proposed an innovative solution to a persistent phenomenon [Rumore, 2016; Howton et al., 2020; Lopez, et al., 2024] in higher education: student disengagement with traditional course syllabi. We hypothesized that presenting syllabi as musical performances, particularly those enhanced by emotionally expressive AI avatars, could significantly improve student attention, comprehension, and retention of critical course information [Zhang et al., 2024; Mycka et al., 2025]. Music has long been recognized as a powerful mnemonic device and emotional catalyst. When integrated with pedagogical content, music can enhance memory encoding and recall, making academic material more engaging, accessible, and enjoyable [Wallace, 1994; Calvert et al., 1993; Moreno et al., 2000].

To implement our approach, we built upon HeyGem[2], an open-source AI singing-avatar project developed by Duix.com. To improve accessibility, we also provided a streamlined Google Colab project, a browser-based platform that lets users run the workflow online rather than install and configure the full system on a local machine. We used Python because it is the standard language for many AI tools and supports the libraries required by HeyGem. In practice, this means that users can execute the avatar-generation pipeline through a simpler web-based interface, reducing both hardware and software barriers. The developed approach allowed users to input text or audio along with a reference video and then generate lifelike singing performances using digital human models powered by deep learning techniques. Here, the reference video is a short video clip of a person or avatar speaking naturally. The reference video serves as the visual template for the output: it provides the character's appearance, facial features, head pose, and general motion style. The system then uses the provided audio track as the new speech or singing content and combines it with the reference video to generate a singing-avatar video. At this stage, the model performs lip synchronization, facial animation, and expression alignment so that the person or avatar in the video appears to speak or sing the supplied audio in a natural and visually coherent way. Additionally, we leveraged Suno AI[3] to transform textual syllabi into structured songs or lyrical narratives. The resulting technology enables educators to transform traditional textual syllabi into fully produced songs performed by virtual avatars, easily shareable via video platforms or directly embeddable LMS like Canvas.

In this paper, we present the concept, implementation, and pedagogical rationale of AI-generated singing syllabi as a design-oriented case study. And we focus on how syllabus content can be reauthored into AI-generated musical and avatar-based media, what design decisions arise during that process, and what practical lessons the workflow offers for instructors interested in more engaging course introductions. The paper mainly contributes an implementation case, a reproducible workflow, and a reflective discussion of opportunities and constraints.

[2] HeyGem is an open-source AI avatar framework alternative to Heygen. HeyGem enables users to generate digital human videos
from text or audio, including synchronized lip movement and facial animation. For more details, see https://github.com/duixcom/Duix.Heygem and https://github.com/Caladog/HeyGem
[3] Suno AI is an artificial intelligence tool that creates full songs from text prompts. For more details, see https://suno.com

## 2. Related Work

In this section, we review existing literature across three intersecting domains: student engagement strategies, music-based learning, and AI-generated educational media. These areas collectively inform the foundation and motivation for our proposed approach, highlighting prior efforts to enhance learner attention, improve memory retention through music, and apply AI in educational contents/topics delivery.

### 2.1 Educational Engagement Strategies

Student engagement is consistently associated with academic achievement and is widely recognized as an important predictor of student success [Wong et al., 2024]. Traditional methods of improving engagement mainly include active learning techniques, flipped classrooms, gamification, multimodal instruction, etc. [Fredricks et al., 2004]. In particular, syllabus engagement has drawn scholarly attention, with research indicating that students are more likely to engage with and retain syllabus content when it is presented in an interactive or personalized format [Ludy et al., 2016]. Innovative formats, such as graphic syllabi, syllabus quizzes, and interactive multimedia introductions, have been proposed to make syllabi more engaging, accessible, and memorable for students [Hanna et al., 2024]. However, the integration of music and AI as a tool to deliver interactive and attractive syllabi remains a largely unexplored frontier [Kim et al., 2024; Shojaei et al., 2025].

### 2.2 Music and Memory in Learning

The relationship between music and memory has been widely studied across cognitive psychology and educational neuroscience. Music, particularly through rhythm and melody, has been shown to enhance verbal memory, reduce cognitive load, and facilitate recall [Wallace, 1994; Calvert et al., 1993]. These effects are especially pronounced when learners are exposed to musical mnemonics or lyrics that structure complex information. While educational songs have traditionally been employed in early childhood and language instruction, they are increasingly being used in higher education [Patel et al., 2010]. For instance, singing has been successfully applied to teaching anatomy, legal terminology, and even mathematics [Yuan et al., 2024]. Building upon this foundation, integrating music into syllabus delivery uses melody not only as a memory aid but also as a tool to emotionally prime students, fostering a stronger connection with course content.

### 2.3 AI in Educational Media

AI is increasingly transforming educational media through personalization, automation, and creative expression [Zhu et al. 2025; Black et al., 2025; Liu et al., 2025]. Advances in deep learning and generative adversarial networks[4] have enabled AI avatars and text-to-music tools to drive applications such as virtual tutors, explainer videos, and interactive storytelling [Holstein et al., 2020]. AI Modern projects, including Ditto [Li et al., 2024], MuseTalk [Zhang et al., 2024], and HeyGem, now support the creation of highly realistic AI characters capable of speaking, singing, and emoting using input scripts and melodies. These tools offer new possibilities for scalable and expressive educational media creation while reducing production costs; however, their effectiveness may depend not only on technical quality but also on learner comfort, perception, and the context in which such avatars are used. Though research has explored AI-generated speech for tutoring and language practice [Chen et al., 2021], little attention has been given to AI-generated singing as a pedagogical medium, particularly for foundational course elements like syllabi.

[4] Deep learning is a branch of machine learning that uses multi-layer neural networks to learn patterns from large datasets; Generative adversarial networks are deep learning models in which one network generates content and another evaluates it, helping produce realistic synthetic data.

In this paper, we proposed a novel solution: using AI-generated singing avatars to perform course syllabi. This approach leveraged the emotional and mnemonic power of music alongside cutting-edge AI avatar synthesis tools to create a syllabus format that is both pedagogically grounded and technologically innovative.

# 3. Case Context and Design Process

In this paper, we employ a design-oriented case-study methodology to develop and document AI-generated singing syllabi aimed at enhancing the presentation of foundational course information. The workflow is structured into four key stages: (1) syllabus selection and lyrical adaptation, (2) audio and music generation, (3) video synthesis, and (4) virtual avatar performance. Below, we describe the case context, clarify the workflow, and surface the design choices and tradeoffs that emerged during implementation.

## 3.1 Case Context and Syllabus Reauthoring

For this pilot implementation, we selected the syllabus for *Introduction to Computer Science*, a foundational course typically offered during the first year of undergraduate computer science curricula. The syllabus includes essential course components such as learning objectives, weekly topics, assessment criteria, scoring guidelines, grading policies, academic integrity expectations, and instructor contact information.

To transform the syllabus into a musical format, we adapted the original textual content into a lyrical script structured around a clear verse-chorus framework. The sections of the syllabus were rewritten into rhythmic, rhyming lines, carefully balancing informational clarity with musicality to maximize student engagement. To support the adaptation process, we initially utilized ChatGPT[5] to generate preliminary lyrical phrases, creative cues, and rhythmic suggestions. These initial outputs were subsequently refined manually to ensure clarity, accuracy, and musical coherence. For instance, Table 1 presents how the original syllabus grading scheme, including the assessment categories and their corresponding point allocations, was creatively transformed into a lyrical excerpt.

[5] https://chatgpt.com/

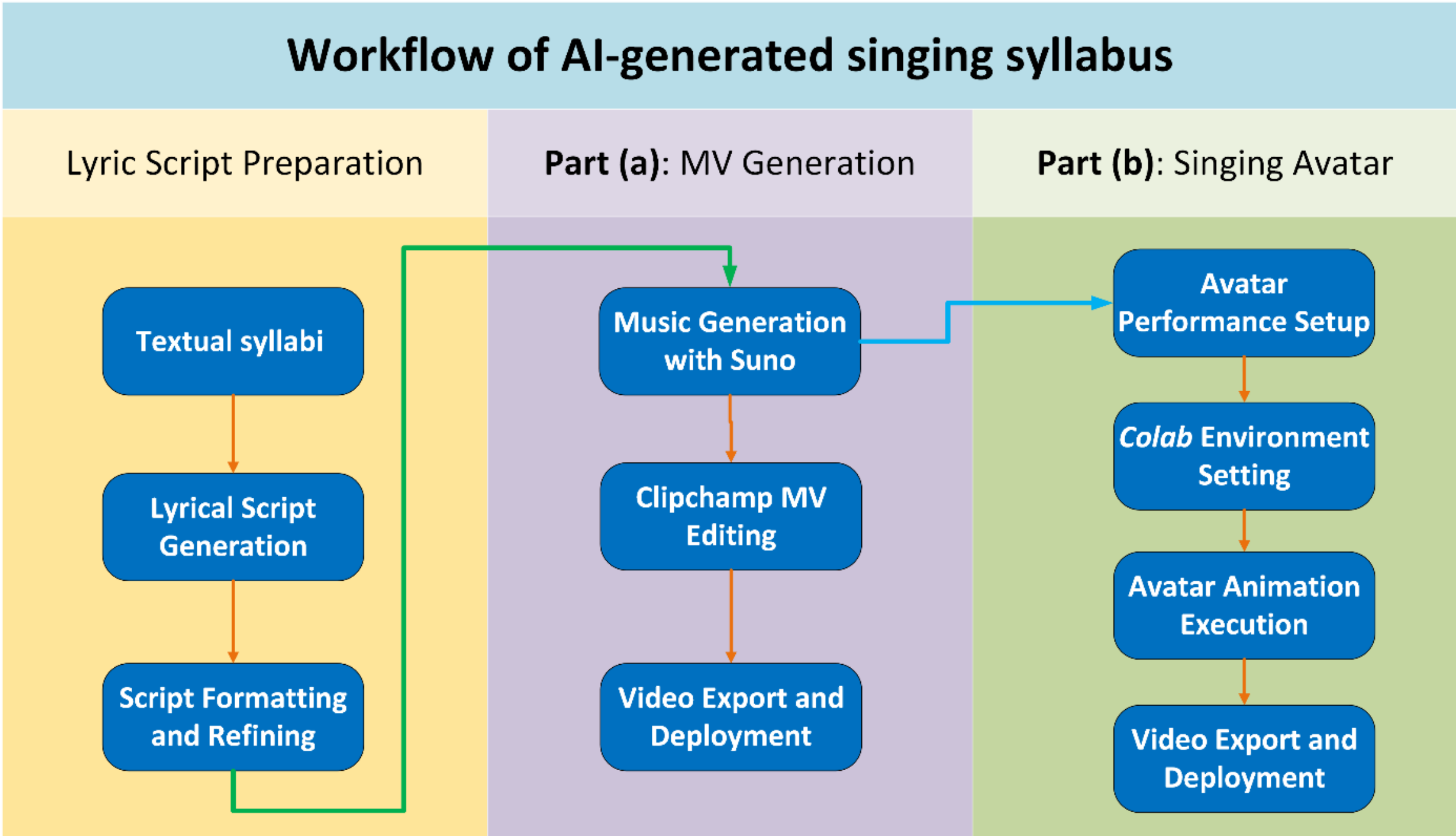

Figure 2. Workflow of AI-Generated Singing Syllabus. Part (a) illustrates the AI Singing and Video Synthesis pipeline. Part (b) presents the Avatar-Enhanced Singing extension by integrating HeyGem for more expressive delivery.

| Value (pts) | Description |
|---|---|
| 50 | Labs, Assignments |
| 20 | Project |
| 10 | Class Attendance |
| 20 | Exam |
| 20 | Challenges |

*Labs and assignments—fifty points to earn,*

*Projects are twenty—show what you've learned.*

*Attendance gives ten—so be here each day,*

*Exams are twenty—prove what you can say.*

*Challenges add bonus—up to twenty more,*

*Altogether, one-twenty's the score to explore!*

***Table 1. An Example of Transformed Lyrics from the Syllabus Scoring Standard.***

The lyrical structure was carefully designed to ensure both accuracy of information and musical coherence.

## 3.2 Audio and Video Prototyping

To bring the lyrics to life, we employed Suno AI , a generative music model capable of transforming text input into high-quality musical compositions in MP3 or WAV format. The genre and mood of the generated music were selected to support the overall instructional tone of the course, favouring a light, upbeat, and welcoming style appropriate for first-year undergraduate students. In this pilot study, the intention was not to establish a formal mapping between musical features and specific learning outcomes, but rather to present the syllabus in a way that felt approachable, attention-grabbing, and easier to remember. This design was motivated by the broader goal of increasing student engagement with syllabus content through a more inviting and emotionally resonant format.

Once the song was generated, we used Microsoft Clipchamp[6] to produce a basic music video (MV) accompanying the audio (For more details, see https://github.com/xinxingwu-uk/SSVA). Visual elements, such as course-relevant icons, thematic illustrations, and animated text lyrics, were incorporated and synchronized with the music to promote clear communication, support recognition of key information, and maintain an engaging but uncluttered presentation [Nielsen, 2024]. In this way, the audio-visual design aimed to reinforce the lyrical content while providing a visually coherent, accessible, and welcoming introduction to the course.

## 3.3 Avatar Extension and Design Rationale

In this extended component of the project, we incorporate virtual avatar performance into the AI-generated singing syllabus using HeyGem, a deep learning-based avatar synthesis framework. To support this integration, we developed a streamlined implementation tailored to our educational use case. This setup allows the generated syllabus song to be presented through a digital avatar, thereby combining audio delivery with synchronized visual performance. Within the present case-study framing, the avatar component is treated as a design extension that illustrates how the workflow can move beyond a lyric video toward a more personalized multimedia course introduction.

The addition of visual personalization introduces expressive facial cues and human-like behaviours that complement the musical delivery. From a design perspective, avatar enhancement offers a way to humanize AI-generated media, support social presence, and create a stronger sense of welcome at the beginning of a course. At the same time, it raises practical questions about tone, realism, accessibility, and instructor comfort, all of which should be considered when adapting this approach in other instructional settings.

# 4. Reproducible Workflow and Technical Implementation

To improve reproducibility, we organize the implementation into two parts: (1) the core singing-syllabus generation pipeline and (2) an optional browser-based interaction extension. The core pipeline is the main contribution used to transform a traditional text syllabus into an AI-generated singing video with avatar performance. The browser-based air-gesture system is an additional deployment interface and is not required for generating the singing syllabus itself.

## 4.1 Core Pipeline: Inputs, Processing Stages, and Output

The proposed system takes four main inputs: (a) the original text-based syllabus, (b) a refined lyric script adapted from the syllabus, (c) a generated audio file in MP3 or WAV format, and (d) a reference avatar video. The final output is an MP4 video in which a virtual avatar performs the syllabus content in sung form.

Building upon the methodological steps described in above, we summarize the complete workflow for transforming a traditional course syllabus into a multimedia AI-driven presentation as the following 7 stages:

[6] https://clipchamp.com/

**Step 1: Lyrical Script Generation**. The original syllabus is first rewritten into a lyrical script. ChatGPT is used only to generate an initial draft, which follows standard musical structures, including verses, choruses, and bridges, to facilitate effective musical adaptation.

**Step 2: Script Formatting and Refinement**. The initial AI-generated lyrical script is saved in plain text format with clear section markers. Then, this script is manually reviewed and lightly revised to improve informational clarity, accuracy, musical coherence, and rhythm before proceeding to the later audio and visual generation stages.

**Step 3: Music Generation with Suno AI**. The finalized lyrics are then provided to Suno AI to generate a complete sung audio track. The generated output is saved as an MP3 or WAV file for the next stage.

**Step 4: Input Preparation for Avatar Synthesis**. The generated song file and the selected reference avatar video are uploaded to the designated project folder on Google Drive. In the current implementation, the reference video may be obtained from multiple sources, such as short instructor-recorded videos, AI-generated avatar templates, or open-source example videos adapted for experimental use. These files serve as the audio and visual inputs to the avatar-generation pipeline.

**Step 5: Google Colab Environment Setting**. To simplify access and reduce hardware requirements, the avatar synthesis process is executed in Google Colab. In this environment (Python3.8), Google Drive is mounted, required libraries are installed, and the runtime is configured with GPU (A100) acceleration. Detailed setup instructions are provided in our GitHub implementation materials[7] .

**Step 6: Avatar Video Generation**. Using the customized HeyGem implementation, the system processes the input audio and reference video to generate a singing avatar video. At this stage, the model performs lip synchronization, facial animation, and expression alignment precisely with the provided audio track.

**Step 7: Export and Deployment**. The generated video is exported as an MP4 file (For more details, see Example on GitHub https://github.com/xinxingwu-uk/SSVA/ ), and can then be embedded in Canvas, uploaded to video platforms, or distributed through other LMS environments.

## 4.2 Runtime Environment and Execution

To support accessibility and reproducibility, we implemented the avatar-synthesis stage in Google Colab rather than requiring local GPU installation. The environment includes Python 3.8, A100 GPU acceleration, mounted Google Drive storage, and required dependencies such as PyTorch and ffmpeg. After the input files are prepared, the main script is executed by specifying the audio path and the avatar-video path. The resulting avatar singing video is automatically saved to the designated output folder in Google Drive.

## 4.3 Optional Browser-Based Interaction Extension - Air-Gesture "Video-in-Air" Playback Demo

As a lightweight extension for classroom deployment, we developed a browser-based prototype that allows users to control playback of the syllabus video using mid-air hand gestures[8]. This component is not part of the core singing-avatar generation pipeline; rather, it serves as an optional playback interface for videos produced by the main workflow, such as a sung-syllabus MV or an avatar-enhanced performance. The demo renders a live webcam background and overlays a floating video panel onto an HTML5 canvas, enabling an AR-style viewing experience without requiring a headset or native app installation. Additional examples and extensions are provided in Part 4 on our GitHub page https://github.com/xinxingwu-uk/SSVA.

[7] https://github.com/xinxingwu-uk/Colab_Implementation-HeyGem/blob/main/Audio2Video.ipynb

**Implementation details**: We use MediaPipe (MediaPipe is an open-source machine learning framework developed by Google for building cross-platform, real-time multimedia processing applications. It provides a suite of libraries and tools that enable developers to quickly apply AI and machine learning (ML) techniques to video, audio, and image data). Hands for real-time hand landmark tracking in the browser and map simple, robust gestures to media controls. A thumb-index pinch gesture toggles play/pause, while a two-hand interaction supports moving, scaling, and rotating the floating video panel. Because modern browsers restrict autoplay with sound, the demo includes an explicit "Enable Sound" interaction to unlock audio playback on first use.

**Pedagogical use**: In an LMS or course website, students can open the demo and interactively control short syllabus micro-videos (e.g., grading policy, weekly topics, integrity policy) on demand. This interaction layer complements the singing-avatar pipeline by offering direct, embodied control and may further improve attention and recall through active engagement.

Because this browser-based interface is intended only as a classroom deployment enhancement, it should be understood as an optional extension rather than part of the core video-generation pipeline.

# 5. Case Study Insights, Design Principles, and Limitations

Because this version is positioned as a case study rather than an evaluation, this section focuses on the design insights, pedagogical implications, implementation constraints, and future research directions that emerge from the workflow.

## 5.1 Design Insights and Practical Guidance

In this paper, first, the proposed approach reframes the syllabus as a communication artifact whose form matters, not merely as a compliance document; Second, it shows how music, narrative structure, and multimodal presentation can be used to foreground key course information that students often ignore in dense text; Third, it illustrates how generative AI tools can support instructional creativity when they are paired with human revision, careful checking, and course-specific judgment.

A further contribution of this work is practical rather than statistical: it offers a reproducible workflow that instructors can adapt to their own contexts. Several design principles emerged from the case. Accuracy should take priority over rhyme. Important policies should be chunked into short, memorable segments rather than compressed into a single dense verse. Visuals should reinforce, not distract from, the message. The emotional tone should invite students into the course without trivializing institutional expectations.

[8] Additional demonstrations are available at the bottom of the project page, including Extension 1 (Demo of Video Air Interaction) and Extension 2 (Demo of Enhanced Air Interaction): https://github.com/xinxingwu-uk/SSVA/.

Finally, the sung syllabus should remain a supplement to the official written syllabus, not a replacement for it.

## 5.2 Limitations, Ethics, and Future Empirical Work

Several limitations and ethical considerations should be acknowledged when interpreting this work and considering future adoption.

### Accuracy and fidelity

When syllabus content is rewritten into lyrics, there is always a risk that precision may be reduced for the sake of rhythm, rhyme, or brevity. Instructors who adopt this approach should therefore validate every transformed line against the official syllabus and preserve a clear link back to the formal text.

Accessibility and institutional alignment: Not all students may prefer, benefit from, or have reliable access to audio-video materials. Captions, transcripts, and the original syllabus document should remain available. Instructors should also ensure that the multimedia version aligns with institutional policy, tone, and course communication norms.

Ethics and future empirical work: This case-study version does not include or analyze student survey data, course-evaluation data, or any other human-subject data. Its purpose is to document the workflow and pedagogical rationale only. Future research can build on this case through ethically approved studies that examine comprehension, recall, engagement, and student perceptions across different syllabus media formats.

## 6. Conclusion

AI is currently employed in diverse fields, such as biology [Wu et al. 2022a; Wu et al. 2022b], psychology [Naddaf, 2025], education [Liu et al., 2025], physics [Wetzel et al., 2025], etc. In this paper, however, we focus specifically on its use for syllabus communication and course introduction. We documented a design-oriented case in which a conventional syllabus was reauthored into lyrics, transformed into music, and extended into video and avatar-based media.

This case study shows that the value of the project lies not only in novelty, but also in its practical workflow and pedagogical positioning. By combining generative AI tools with human editing and instructional judgment, educators can create syllabus media that are more welcoming, memorable, and aligned with contemporary digital communication practices.

Although this paper currently does not claim empirical effects on students, it establishes a concrete implementation pathway and a reflective framework for future work. As AI media tools continue to evolve, design-based case studies such as this can help the field think more carefully about when, why, and how multimedia syllabus communication should be developed and evaluated.

# Appendix 1. Transformed Lyrics from the Syllabus

Welcome to Intro to Computer Science

**[Verse 1]**

Welcome to Intro to Computer Science,

Where data meets logic, and concepts combine.

I'm Wu, your teacher—find me at Starks 215D,

Office hours posted on Canvas, come see me.

Drop by anytime, I'm always near,

My accent's not native, so ask if unclear.

**[Chorus]**

We're gonna learn, create, and grow,

Explore the field where the future will go.

From SQL to AI, Python to the web,

Building foundations, step by step.

Our journey is basic, but vast in scope,

Across computer science, with curiosity and hope.

Sixteen weeks to unlock your mind,

With hard work and passion, success you'll find!

**[Verse 2]**

Each week builds on what we've done before,

Databases, modeling, and so much more.

HTML and CSS to design with care,
JavaScript adds the interactive flair.
Networking, algorithms, and AI in sight,
This course lays the groundwork for tech's flight.

**[Chorus]**

We're gonna learn, create, and grow,
Explore the field where the future will go.
From SQL to AI, Python to the web,
Building foundations, step by step.
Our journey is basic, but vast in scope,
Across computer science, with curiosity and hope.
Sixteen weeks to unlock your mind,
With hard work and passion, success you'll find!

**[Bridge 1: Submission Requirements]**

Submit your work before the deadlines appear,
Show your progress, make everything clear.
Projects and challenges, rise to the test,
Push your limits, always give your best.
Final presentation—mark that date,
April 28th, 10:30 to 12:30—don't be late!

**[Bridge 2: Scoring Breakdown]**

Labs and assignments—fifty points to earn,
Projects are twenty—show what you've learned.
Attendance gives ten—so be here each day,
Exams are twenty—prove what you can say.
Challenges add bonus—up to twenty more,
Altogether, one-twenty's the score to explore!

**[Chorus]**

We're gonna learn, create, and grow,
Explore the field where the future will go.
From SQL to AI, Python to the web,

Building foundations, step by step.
Our journey is basic, but vast in scope,
Across computer science, with curiosity and hope.
Sixteen weeks to unlock your mind,
With hard work and passion, success you'll find!

**[Bridge 3: Grading Scale]**

An A for excellence—ninety or more,
A B for strong work—eighty's your floor.
C is acceptable—seventy will do,
D means improvement—it's all up to you.
Below sixty is failing, but don't despair,
With effort and time, success is there!

**[Outro]**

We'll use Brookshear's book, the 13th edition,
A guide to computer science, your mission.
Code, design, and innovate—mark the date:
April 28th, your final awaits!
Let's make it great in Intro to Computer Science,
Where your tech journey starts, and dreams advance.